\documentclass[twocolumn]{aastex631}

\usepackage{soul}
\usepackage{natbib}
\usepackage{tablefootnote}

\usepackage{CJK}
\usepackage{amsmath}

\begin{document}

\title{The Relation Between AGN and Host Galaxy Properties in the JWST Era: \\
II. The merger-driven evolution of Seyferts at Cosmic Noon}

\correspondingauthor{Nina Bonaventura}
\email{nbonaventura@arizona.edu}

\author[0000-0001-8470-7094]{Nina Bonaventura}
\affiliation{Steward Observatory, University of Arizona, 933 North Cherry Avenue, Tucson, AZ 85721, USA}

\author[0000-0002-6221-1829]{Jianwei Lyu (\begin{CJK}{UTF8}{gbsn}吕建伟\end{CJK})}
\affiliation{Steward Observatory, University of Arizona,
933 North Cherry Avenue, Tucson, AZ 85721, USA}

\author[0000-0003-2303-6519]{George H. Rieke}
\affiliation{Steward Observatory, University of Arizona,
933 North Cherry Avenue, Tucson, AZ 85721, USA}

\author[0000-0002-8651-9879]{Andrew J. Bunker} 
\affiliation{Department of Physics, University of Oxford, Denys Wilkinson Building, Keble Road, Oxford OX1 3RH, UK}

\author[0000-0002-4201-7367]{Chris J. Willott}
\affil{NRC Herzberg, 5071 West Saanich Rd, Victoria, BC V9E 2E7, Canada}

\author[0000-0001-9262-9997]{Christopher N. A. Willmer}
\affiliation{Steward Observatory, University of Arizona, 933 North Cherry Avenue, Tucson, AZ 85721, USA}

\begin{abstract}

 In Paper I, we exploited the unsurpassed resolution and depth of JWST/NIRCam imagery to investigate the relationship between AGN and host-galaxy properties in the JWST era, finding a correlation between the level of spatial disturbance (as measured by shape asymmetry, $A_S$) and obscuration ($N_H$). Here in Paper II, we report an expansion of our X-ray and infrared analysis of Seyfert-luminosity host galaxies with four additional metrics to the single-metric morphology analysis of Paper I, as well as new samples of inactive control galaxies. This expanded study of one of the largest and most complete, multi-wavelength samples of AGN detected at $0.6<z<2.4$ in the GOODS-South and North fields, confirms that mergers surprisingly play a significant role in obscured, sub-quasar AGN host galaxies. Additionally, the pattern of morphological disturbances observed amongst the X-ray- and mid-IR-selected AGN suggests that these represent different phases of AGN evolution tied to a major-merger timeline, as opposed to distinct populations of AGN. These results indicate that mergers are important in triggering sub-quasar AGN at these redshifts.
 
\end{abstract}

\section{Introduction}

The role of galaxy mergers in triggering and feeding active galactic nuclei (AGN) is a central prediction of a number of theoretical simulations, but has been difficult to demonstrate observationally. The numerous investigations into this suspected AGN-merger connection have not converged on a solid result, as discussed, for example, in the review by \citet{Villforth2023} (see also e.g., \citet[][]{Zhao2022, Omori2025}). However, as studies have moved from selections favoring lightly obscured AGN (through optical colors, broad emission lines, and X-ray detection) to those inclusive of heavier obscuration (typically sources faint in the optical and/or X-ray), this picture has changed; nearly every study of {\it obscured} AGN has found that they tend to  reside in galaxy mergers \citep{Kocevski2015,  Ricci2017, Weston2017, Donley2018, Goulding2018, Ellison2019, Peifle2019, Gao2020, Barrows2023, Kim2024, Euclid2025, Bonaventura2025}.

In general, these results have been confined to very luminous AGN, in part because Wide-field Infrared Survey Explorer (WISE) \citep{Wright2010} infrared data have played a key role in identifying obscured AGN: the Spectral Energy Distributions (SEDs) of WISE-detected AGN, by definition, require a relatively high AGN luminosity to appear against the competing emission from the host galaxy in this portion of the SED \citep{Hickox2018}. 

The theoretical expectation has been that lower-luminosity AGN, unlike quasars, can passively grow through stochastic processes without being associated with mergers \citep[e.g.,][and references therein]{Hopkins2006, Hopkins2009, Mcalpine2018,  Weigel2018}. However, before the advent of JWST, it was difficult to investigate the role of mergers in the triggering and evolution of obscured, lower-luminosity AGN, as there were no mid-infrared surveys with adequate sensitivity and wavelength coverage to yield a significant sample of such AGN. Furthermore, it was difficult to obtain detailed rest-optical (observed near-infrared) morphologies of high-redshift galaxies hosting moderate-luminosity AGN, due to limited observing capability in this wavelength range, and the inability of JWST's predecessors to overcome the surface brightness dimming caused by cosmological expansion \citep[e.g.,][]{Shi2009}. This issue drove studies to probe the frequencies and characteristics of obvious mergers, where two galaxies or two nuclei appear, rather than morphological indications of disturbance in single galaxies (i.e., candidates for merger remnants) \citep{Shah2020}. 

Consequently, it is possible that late post-coalescent merger remnants have been frequently overlooked and misclassified as non-mergers, and therefore excluded from merger studies, given that the telltale signatures of post-coalescence, such as tidal tails, wisps, and cusps around the galaxy edge, would typically have been too faint to be observed. 

In Paper I, the superior sensitivity and resolution of JWST/NIRCam enabled the discovery of strong galaxy-scale spatial disturbances in a significant fraction of our AGN sample at $z>~0.5$, which was unexpected given that sub-quasar AGN are thought to be triggered predominantly by non-merger mechanisms. This finding suggests that merger signatures persist in remnants for longer than previously thought \citep[e.g.,][]{Ellison2025}, and that Seyfert activity is enhanced significantly  with the pervasive merging found to occur at Cosmic Noon \citep[e.g.,][]{Conselice2022, Puskas2025}.

Another key component of our analysis that allowed for the detection of significant evidence of prior merging activity in our AGN sample, was our use of the shape asymmetry ($A_S$) parameter in diagnosing merging morphologies in imaging data (\citealp{Pawlik2016, Nevin2019}). This non-parametric morphology indicator provides a clean and unbiased measure of the level of spatial disturbance along the edges of a galaxy by using the binary detection image corresponding to its flux image -- meaning it is uninfluenced by brightness asymmetries unrelated to merging activity, unlike the classic asymmetry parameter, \(A\). 

In the current study, in addition to comparing the AGN host morphologies to a new sample of 784 matched inactive (non-AGN) control galaxies, we build upon the results of Paper I by continuing our analysis of 425 AGN selected at X-ray and mid-IR wavelengths from $z\sim0.5$ to $z\sim4$, in the following ways:

1) We expand the AGN host-galaxy morphology analysis to encompass four additional metrics to the shape asymmetry reported in Paper I, allowing us to distinguish disturbances caused by major versus minor mergers, and therefore assign a merger type to each AGN in the sample. This additional information allows us to determine if merging activity in Seyferts is significant to their evolution and SMBH growth, as would be indicated by significant evidence of major-merging activity.

2) We investigate if the SED-derived properties of the mid-IR and X-ray AGN host galaxies support the results of the morphology analysis: that these two subsamples are not distinct AGN classes, but likely different phases of an AGN evolutionary sequence, where the mid-IR/X-ray-faint phase chronologically precedes the X-ray-bright phase and occurs closer in time to the statistically likely major merger that drives their evolution.

3) We investigate if the conclusion we draw from the combined morphology and SED analysis of this representative sample of Seyferts holds up in comparison to control samples of non-AGN galaxies matched on stellar mass, redshift, specific star formation rate (sSFR), and selection wavelength.

As in Paper I, we adopt the cosmology of the studies from which we draw our AGN sample, namely \citet{Lyu2022} and \citet{Lyu2024}, who adopt a slighty different cosmology: \citet{Lyu2022} assumes $\Omega_m=0.27$ and $H_0 = 71$ km~s$^{-1}$Mpc$^{-1}$, and \citet{Lyu2024} assumes $\Omega _m = 0.287$ and $H_0=69.3$ km~s$^{-1}$Mpc$^{-1}$. This minor difference does not affect the conclusions of our study.

\section{Samples and Data}
In the present study we extend our analysis of the primary AGN sample considered in Paper I, which is comprised of two subsamples: (1) 243 X-ray-selected AGN analyzed in \citet{Lyu2022} (GOODS-S) and J Lyu (private communication, 2025, GOODS-N); and (2) 182 mid-infrared(mid-IR)-selected AGN analyzed in \citet{Lyu2024} (GOODS-S), for a total of 425 AGN.

In Paper I, in addition to the full primary sample, we analyzed a subset consisting of only those X-ray and mid-IR AGN matched on various source properties, to rigorously quantify a relationship between AGN obscuration and spatial disturbance. Here, we consider the full, unmatched primary sample, at the redshifts where it is complete ($0.5<z<3.8$), to uncover a suspected relationship between the X-ray and mid-IR AGN. 

\subsection{X-ray AGN Sample}
The X-ray AGN sample consists of 243 lightly-to-highly obscured \textit{Chandra}-detected sources lying at redshifts $0.5<z<3.8$, with $N_H$ values in the range $10^{22}$-$10^{24}$ $cm^{-2}$, as derived from hardness ratios \citep{Liu2017,Luo2017}. Their stellar masses\footnote{The stellar masses inferred are based on the stellar population synthesis from the \textit{Prospector} code, and the AGN component has been modeled by a set of well-tested, semi-empirical templates (see \citet{Lyu2024} for details). We note that the vast majority of the galaxies in our sample have highly obscured AGN, therefore AGN contamination is not expected to impact their stellar mass estimates, as demonstrated by \citet{Ciesla2015}.} ($M_*$) and bolometric luminosities ($L_{bol}$) are estimated from the SED-fitting procedure described in \citet{Lyu2022}, and lie in the range $10^9-10^{11.5}$ $M_\odot$ and $10^{41}-10^{46}$ erg $s^{-1}$, respectively. A description of the full set of properties relevant to this study is contained in Paper I.

\subsection{Mid-IR AGN Sample}
The mid-IR AGN sample was relatively recently identified with the Systematic Mid-infrared Instrument Legacy Extragalactic Survey (SMILES) \citep{Lyu2024}, and consists of many highly obscured (X-ray-faint) systems, $80\%$ of which were previously undetected by extensive pre-JWST AGN searches in GOODS-S \citep{Lyu2024}. We include 182 AGN in our analysis with the following SED-derived properties: $10^9 \le M_*/M_{\odot} \le 10^{11.4}$, $10^{42} \le L_{bol} \le 10^{46}$, and $0.5<z<3.8$. These sources were classified as AGN through an SED analysis making use of twenty-seven photometric bands, including five HST/ACS bands (0.44 - 0.9 $\mu$m), fourteen JWST/NIRCam bands (0.9 - 4.4 $\mu$m), and eight JWST/MIRI bands (5.6 - 25.5 $\mu$m) \citep{Lyu2024}. A description of the full set of properties relevant to this study is contained in Paper I.\footnote{We note that our mid-IR AGN sample is fundamentally different from previous ``IR AGN" samples that were selected on the basis of \textit{Spitzer}/IRAC or WISE infrared colors: as is shown in \citet{Donley2012}, these color-selection methods reveal AGN SEDs described by a power law model, which means that at significant redshift, the power law shifts into the near-infrared. Thus, IR AGN selected according to this method are {\it not} heavily embedded, but instead are moderately obscured, Type-1 AGN. In comparison, our sample was detected on account of the significantly improved sensitivity and continuous spectral coverage of JWST/MIRI.}

\subsection{Inactive Control Samples}

\begin{figure}
\centering
\includegraphics[width=1\linewidth]{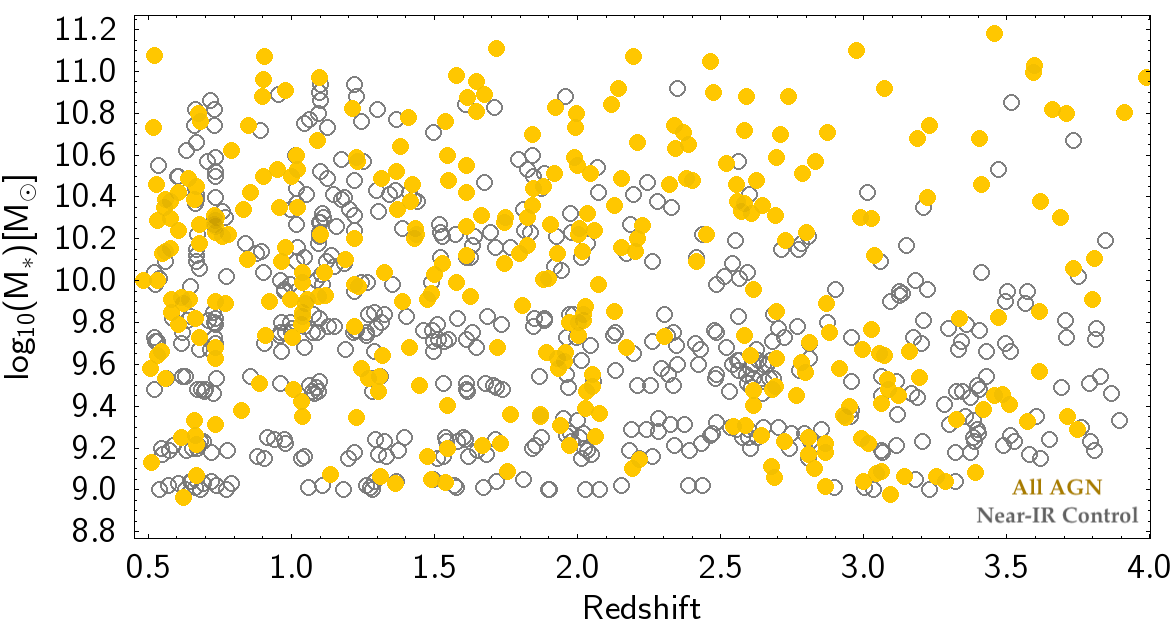}  
\centering
\includegraphics[width=1\linewidth]{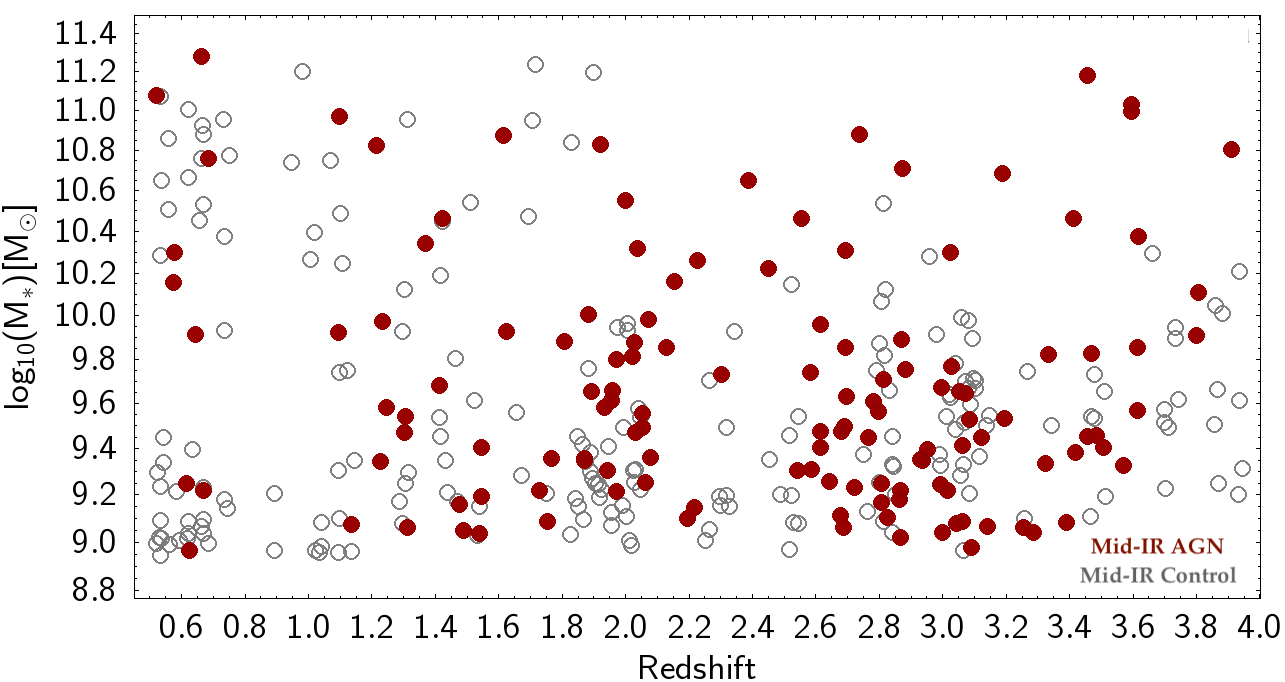}  
\centering
\includegraphics[width=1\linewidth]{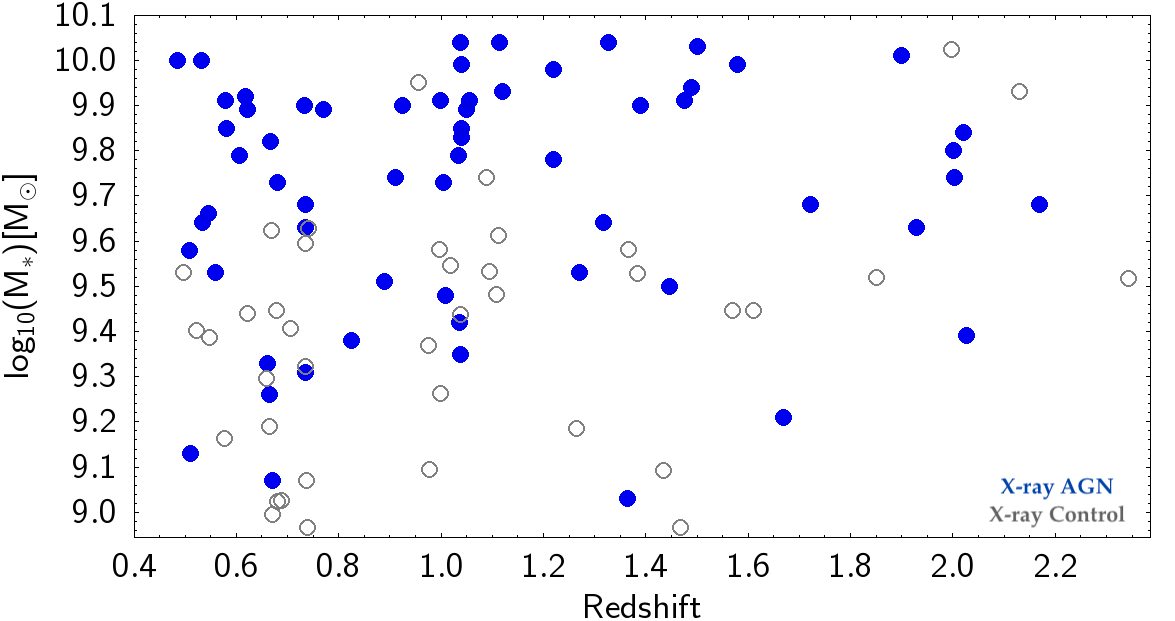}  
\caption{The distributions of the stellar masses and redshifts of the inactive control samples matched to the AGN sample. The top figure shows the near-IR-selected control sample from CANDELS with the full primary AGN sample in yellow. The middle figure shows the mid-IR AGN (red) and corresponding MIRI-selected control sample. The bottom figure shows the X-ray AGN (blue) compared to the X-ray-selected control sample.}
\label{fig:control_match}   
\end{figure}

To determine if the morphology patterns we observe in our AGN sample are unique to AGN, and therefore that we are truly uncovering an  evolutionary mechanism in Seyferts at Cosmic Noon, we repeated the full host-galaxy morphology analysis of Paper I on a control sample of inactive galaxies. We collected three different control samples to (1) guard against unsuspected systematic effects that might plague a single sample; (2) be as comprehensive as possible in the property-matching to the different AGN subsamples by matching on selection wavelength; and (3) to maximize the number of inactive controls for better statistics.

\subsubsection{Near-IR-selected Inactive Galaxies}

For the primary inactive control sample, we utilized the stellar-mass catalog for the GOODS-S field detailed in \citet{Santini2014}, which was selected using deep near-IR HST photometry as part of the Cosmic Assembly Near-infrared Deep Extragalactic Legacy Survey (CANDELS) project. These masses were derived by template fitting constrained by $\sim$ 10 photometric bands \citep{Grazian2006}. From this catalog, we extracted sources that were not flagged as AGN or stars; that have stellar masses\footnote{The stellar-mass catalog in \citet{Santini2014}  contains ten different $M_*$ estimates; we utilized the $11 \mathrm{a}_\tau$ method, described in Section 3.2 of that paper. This method utilizes the \citet{Bruzual2003} stellar templates, a Chabrier IMF, and an exponentially decreasing SFH (direct-$\tau$ model).} in the $9<log(M_*/M_{\odot})<11$ range and $SFR<100$ $M_{\odot}/yr$\footnote{Florian et al. (2025, ApJ, accepted) demonstrate insignificant morphological disturbance in this range of SFR.}; and $SNR \ge 30$ at $0.5<z<3.0$ and $SNR \ge 10$ at $3.0<z<4.0$ (to boost the number of sources selected at highest redshifts, where fewer sources were available than at lower redshifts). This resulted in a sample of 529 control galaxies. 

\subsubsection{Mid-IR-selected Inactive Galaxies}

The second inactive control sample was chosen to match the mid-IR AGN subsample, and therefore was drawn from the same MIRI SMILES catalog and analyzed using the identical SED analysis as the AGN sources in the catalog. We required these sources to be classified as non-AGN, with a SFR less than $100$ $M_{\odot}/yr$ (which also limited $L_{IR}<10^{12}$ $L_\odot$), and with stellar masses ($9<log(M_*/M_\odot)<11$) and redshifts ($z<4$) matched to the mid-IR AGN. We found 213 such inactive control galaxies that were included in the JADES NIRCam v1 mosaic image used for morphology analysis. 

\subsubsection{X-ray-selected Inactive Galaxies}

The third inactive control sample was drawn from the X-ray source catalog of \citet{Luo2017}, which contains information on hundreds of X-ray-identified sources within the 3D-HST footprint and therefore overlaps with our X-ray-selected AGN sample in GOODS-S. The list of inactive control galaxies was gathered by first identifying those in the catalog designated as non-AGN, and then further filtering from that list the lingering AGN contaminants using the superior AGN detection methods of \citet{Lyu2022}. Next, we calculated the stellar masses of the non-AGN from their observed $K_s$ magnitudes, and then matched to the X-ray AGN sample on stellar mass and redshift. Finally, we removed those galaxies found to lie outside the footprint of the JADES NIRCam v1 mosaic imagery used for morphological analysis, leaving 42 X-ray-selected inactive control galaxies. While the X-ray source catalog of \citet{Luo2017} does not provide estimates of SFR, \citet{Lehmer2016} find that $96\%$ of the non-AGN in this sample have SFR $<$ 100 M$_\odot$ yr$^{-1}$, and therefore our X-ray AGN and control samples are also matched on SFR.

We note that, unlike for the near-IR and mid-IR control samples, the ranges of both redshift and stellar mass on which the X-ray control sample could be matched to the X-ray AGN are smaller than the full redshift ($0.5<z<4$) and stellar-mass range ($9<log(M_*/M_\odot)<11$) of the primary X-ray AGN sample, as shown in the bottom plot of Figure\ref{fig:control_match}.

\subsection{JWST NIRCam Data}

As in Paper I, for the AGN host-galaxy morphological analysis, we utilize the NIRCam F150W mosaic imagery from version 1 of the JWST Advanced Deep Extragalactic Survey (JADES). This survey is introduced in JWST programs 1180 and 1181 (PI: Eisenstein), and fully described in the JADES Data Release v1 paper \citep{Rieke2023} and JWST Extragalactic Medium-band Survey (JEMS) (JWST PID 1963) paper \citep{Williams2023}. Each mosaic image achieves $0.03''$/pixel resolution and an AB magnitude depth of 29-30 \citep{Rieke2023}. 

\section{Methodology}
\subsection{Characterizing AGN Obscuration}

Our sample is selected on the basis of X-ray and infrared properties and is almost entirely obscured, i.e. of type S1.8 - S2, or Compton-thick, according to the usual identification of obscured AGN as those with $log(N_H) > 22$ \citep[e.g.,][]{Risaliti1999, Padovani2017, Kammoun2020, Annuar2025}. As shown in the line-of-sight (LOS) X-ray absorption distribution in Figure 1 of Paper I, the sample contains very few examples at $log(N_H) < 22$, with most lying in the $22<log(N_H)<24+$ range. This sampling will typically have obscured nuclei in the optical \citep[e.g.,.][]{Burtscher2016, Padovani2017}, leading to low contrast against the stellar population of the galaxy -- a prediction that agrees well with visual inspection of our sample (see the examples in Figure~\ref{fig:LDA1} and in Figure 3 of Paper I). We have therefore conducted the morphological analysis without correcting for a central point source, given that only a very small fraction of the galaxies under study exhibit one. 

\subsubsection{$N_H$ Measurements}
For the AGN well-detected in the X-ray, the column density, $N_H$, is a convenient and widely used indicator of obscuration. However, as it applies only to the column along the line of sight, one must probe whether the possibility of highly non-isotropic obscuration undermines its application in this regard. In other words, if the AGN output can escape unobscured in directions away from our line of sight, it should excite emission lines in ionization cones, i.e. in the narrow line region; if this region is large and only lightly obscured, then it would violate the assumption of isotropic obscuration. 

\citet{Lamassa2010} provide a useful summary in the form of extinction from Balmer decrements for a sample of 40 Type 2 AGN. Only a minority of their sample exhibit orientations close to `edge-on', where extinction might arise from the disk of the galaxy, and these members are similar in behavior to the rest. We convert the Balmer decrements to $A_V$ using the results in \citet{Koyama2015, Gordon2021}, and translate the resulting values of $A_V$ to $N_H$ from \citet{Guver2009}. The average value is $N_H = 8 \times 10^{21}$ cm$^{-2}$, i.e., a column close to the threshold for LOS obscuration in a Type 2 AGN. In other words, the directions away from high obscuration regions encounter, on average, reduced, but still very significant, obscuration.

\subsubsection{Dust Torus Covering Fraction}

Considering the covering fraction of the heavily obscuring material -- i.e., how much of the isotropic emission will encounter a high level of obscuration -- \citet{Ricci2017} have modeled the behavior of sources in the Swift/BAT 70-month AGN Catalog. For those with $N_H > 0.5 \times 10^{22}$ cm$^{-2}$, the average covering fraction is about 65\%. \citet{Lanz2019} also analyzed Swift/BAT measurements, supplemented with data from \textit{Herschel} and \textit{NuStar}; they deduced average covering fractions of $70 - 80\%$, in agreement. Thus, there is a significant minority where the column is less or more than the observed $N_H$, but from the results discussed in Section 3.1.1, this column is by no means negligible.  Of course, the effect of this minority is two-way; our line of sight might intersect the lightly obscured path rather than the heavily obscured one. 

The effect, then, of these two issues together, is to increase the scatter of relations based on obscuration estimated from the LOS $N_H$, but not to preclude them. Therefore, when interpreted in a statistical sense, conclusions drawn from $N_H$ as a metric for the obscuration should be valid.

\subsection{Multi-metric Merger Morphology Analysis}

In Paper I, we relied on the shape asymmetry parameter ($A_S$) to identify merger signatures in imaging data, following the procedures developed by \citet{Nevin2019} (hereafter `N19'), as shown in Figure~\ref{fig:AStrend}. Those authors simulated SDSS \textit{r}-band images of local, major- and minor-merging galaxies, and used \textit{statmorph} \citep{Rodriguez2019} to generate a large statistical sample of their non-parametric morphology measurements; they then fit those parameters together in a linear discriminant analysis (LDA) that proved to robustly diagnose merging morphology in imaging data. From that study, it became apparent that $A_S$ individually outperforms the other metrics in diagnosing merger signatures in imaging data (as was also shown in \citet{Pawlik2016}, where $A_S$ was introduced), and over the longest length of the merger timeline -- only surpassed by the LDA diagnostic that combines all morphology metrics considered in their study. Therefore, $A_S$ dominates the calculations used in N19 to quantitatively classify merger signatures in images. In \citet{Nevin2023} (N23), the authors expand and refine the LDA analysis to robustly distinguish major from minor mergers, as well as to identify merger stage. 

\begin{table*}
\label{morph_metrics}
\centering
\caption{The Linear Discriminant Analysis (LDA) metrics from Table 2 of \citet{Nevin2023} for the classification of host-galaxy merger type and stage, where the four leading coefficients and terms dictating each merger classification are shown, based on the corresponding fit to the full set of galaxy morphology parameters (noting the removal of the 0.5-Gyr post-coalescence classifications, as they show lower performance statistics than the 1.0-Gyr post-coalescence classifications; see Section 4.3 of N23 for details). In our final analysis, we utilized only the combined major/minor merger classifications shown, as the \(Gini\) and/or S$\acute{e}$rsic \(n\) values required for most of the other classification metrics were unusable for a significant fraction of our sample, as discussed in the text.} 
\begin{tabular}{|l|l|l|l|l|}
\hline Classification & Term 1 & Term 2 & Term 3 & Term 4 \\
\hline \textbf{All major mergers} & \(13.9 \pm 1.0 A_s\) & \(-8.0 \pm 0.7 A_s * C\) & \(-5.4 \pm 0.4 A_s * A\) & \(5.1 \pm 0.4 A\) \\
\hline Major, pre-coalescence & \(10.0 \pm 0.6 A_s\) & \(7.5 \pm 0.2 A\) & \(-6.3 \pm 0.2 A_s * A\) & \(-6.1 \pm 0.5 A_s * C\) \\
\hline Major, early stage & \(9.1 \pm 0.4 A_s\) & \(-5.8 \pm 0.4 A_s * C\) & \(5.3 \pm 0.6 C\) & \(4.9 \pm 0.5 A\) \\
\hline Major, late stage & \(-8.9 \pm 0.8 A_s * A\) & \(7.9 \pm 0.4 A_s\) & \(7.2 \pm 0.7\) Gini*A & \(1.2 \pm 0.2 A * S\) \\
\hline Major, post-coalescence (1.0 Gyr) & \(-14.3 \pm 0.9 C * n\) & \(11.7 \pm 1.4 C\) & \(5.9 \pm 0.9\) Gini*n & \(-1.3 \pm 0.2 A_s * M_{20}\) \\
\hline \textbf{All minor mergers} & \(-10.4 \pm 1.9 A_S * C\) & \(8.8 \pm 0.7 A * C\) & \(-7.8 \pm\) 3.3 Gini*S & \(-7.8 \pm 0.6 A\) \\
\hline Minor, pre-coalescence & \(-31.3 \pm 7.7\) Gini*S & \(-28.6 \pm\) 6.0 Gini*n & \(27.4 \pm 5.7 n\) & \(21.0 \pm 2.8 C\) \\
\hline Minor, early stage & \(20.8 \pm 3.6 C\) & \(-20.5 \pm 5.4\) Gini*C & \(-18.0 \pm 2.2 n * M_{20}\) & \(-16.7 \pm 2.2 n * C\) \\
\hline Minor, late stage & \(10.1 \pm 1.4 A_s * C\) & \(-5.3 \pm 1.0 A_s *\) Gini & \(1.9 \pm 0.1 A_s * A\) & - \\
\hline Minor, post-coalescence (1.0 Gyr) & \(2.0 \pm 0.1\) Gini & \(-1.1 \pm 0.1 A * S\) & \(0.6 \pm 0.1 n\) & - \\
\hline
\multicolumn{5}{l}{\footnotesize{\textit{Note}: The morphology predictor values entered should be standardized based on the mean and standard deviation of each classification.}}
\end{tabular}
\end{table*}

As introduced in N19, the imaging component of the LDA framework performs a supervised classification of merging versus non-merging galaxies by identifying the linear combination of morphological features that separates them (equations 1 and 2 of N23), including \(A_S\), the \(Gini\) coefficient, the \(M_{20}\) statistic, Concentration (\(C\)), Asymmetry (\(A\)), Clumpiness (\(S\)), and S$\acute{e}$rsic index (\(n\)):

$$
\begin{aligned}
\mathrm{LD} 1_{\text {major }}= & +13.9 A_s-8.0 C * A_s-5.4 A * A_s+5.1 A \\
& +4.8 C-2.9 \mathrm{Gini} * A_s+0.6 M_{20} * A \\
& +0.4 M_{20} * n+0.4 \mathrm{Gini}-0.6
\\
\\
\mathrm{LD} 1_{\text {minor }}= & -10.4 \mathrm{C} * A_s+8.8 \mathrm{C} * \mathrm{~A}-7.8 \mathrm{Gini} * \mathrm{~S}-7.8 \mathrm{~A} \\
& +6.6 \mathrm{~A}_s+6.5 \mathrm{Gini} * \mathrm{M}_{20}-6.0 \mathrm{M}_{20} * \mathrm{~S} \\
& -5.7 \mathrm{M}_{20} * \mathrm{~A}_s+4.9 \mathrm{~S}-4.4 \mathrm{M}_{20}+3.7 \mathrm{Gini} * \mathrm{C} \\
& -2.9 \mathrm{~S} * \mathrm{n}-1.0 \mathrm{n} * \mathrm{~A}_s-0.2 \mathrm{~A} * \mathrm{~S}-0.7
\end{aligned}
$$

The results of N19 showed that their LDA method, which considers seven different morphology metrics, outperforms popular merger classification methods such as the $Gini-M_{20}$ diagnostic, and the use of $A$ or $A_S$ alone. In N23, the LDA classifier is used to separate merging and non-merging galaxies of varying merger mass ratios (major versus minor) as well as the merger stage (early, late, post-coalescent), in a sample of 1.34 million local SDSS galaxies. 

To expand upon the AGN host-galaxy morphology analysis of Paper I to encompass a broader set of morphology metrics, and therefore determine the merger type and stage indicated by the shape asymmetries of Paper I, we again used \textit{statmorph} to generate the full set of morphology measurements required to conduct the LDA analysis. As in N23, we classified our galaxy sample according to the four leading coefficients and terms for each of the merger type and stage classifications contained in Table 1.

The predictor values (as they are referred to in N19 and N23) we measured for each morphology parameter for the AGN hosts were consistent with the distributions of those measured for the local SDSS galaxies (see Figure 1 of N23), for all but the most heavily obscured systems\footnote{We learned during this analysis that the \(Gini\) coefficient values measured for our Compton-thick (CT) AGN lie outside the parameter space of the SDSS galaxies, indicating a new area of exploration for very heavily obscured examples not included in the N23 study. We also observed for our most-obscured AGN, that the S$\acute{e}$rsic fitting failed because the CT AGN are the most spatially disordered, and therefore we could not in those instances obtain a measure of the S$\acute{e}$rsic index required for a number of the merger classification formulae in N23. In future work, we plan to conduct a similar analysis to N23 to explore this new parameter space for highly obscured galaxies observed in JWST data.}, therefore we directly applied the N23 LDA model to our NIRCam data. In other words, we did not simulate JWST/NIRCam images of galaxies and then refit to this data the LDA model of N23 to tune the parameters accordingly, as we consider their model to be generalizable to NIRCam data: the only instrumental difference we expect to manifest in our NIRCam images is a higher SNR than the simulated SDSS images, despite the higher redshifts we probe (see Fig. 11 of Paper I for a comparison of a sample of our real NIRCam images to their simulated SDSS galaxies at matched rest-frame wavelength). 

Given that the \(Gini\) and/or \(n\) metrics required for refining the classification of merger type into merger stage were not available or valid for a significant fraction of our sample, we utilized only the combined major/minor merger LDA metrics for determining merger type, using the four leading coefficients and terms of Equations 1 and 2 (also shown in Table 1) as in N23. Where the \(Gini\) coefficient values were invalid (for the N23 analysis) for the most obscured AGN ($41\%$ of the sample), we could not classify those galaxies as minor mergers, due to the dependence of the minor-merger classification metric on \(Gini\). However, all strongly classify as major mergers at $98.2\pm5.6\%$, therefore we assume these AGN to be associated with major mergers. 

\begin{figure}

\centering    
\includegraphics[width=0.95\linewidth]{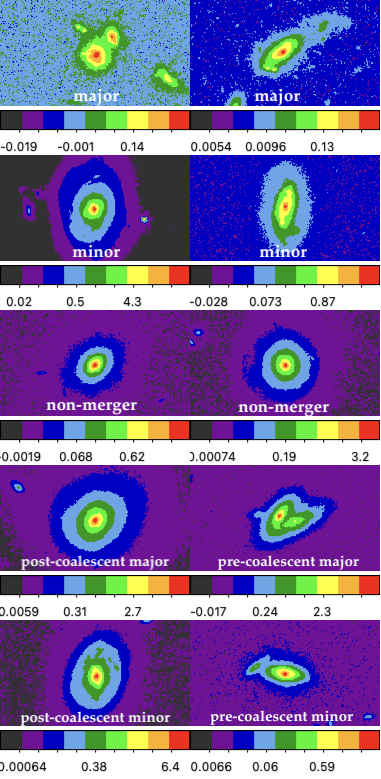}
\caption{Examples of AGN host galaxies that were classified as major, minor, and non-mergers by the combined major/minor merger classifications shown in Table 1, used to diagnose the likely merger stage corresponding to the spatial asymmetries uncovered in Paper I. Additionally, we show examples with the full set of morphology indicators required for a further refinement of the classification to merger stage. These images were cut from the F150W NIRCam v1 mosaics used for morphology characterization in the current paper and Paper I, with pixel intensities colored with SAOImage `aips0' and shown on a log scale.}
\label{fig:LDA1}   
\end{figure}

\begin{figure}

\centering    
\includegraphics[width=1\linewidth]{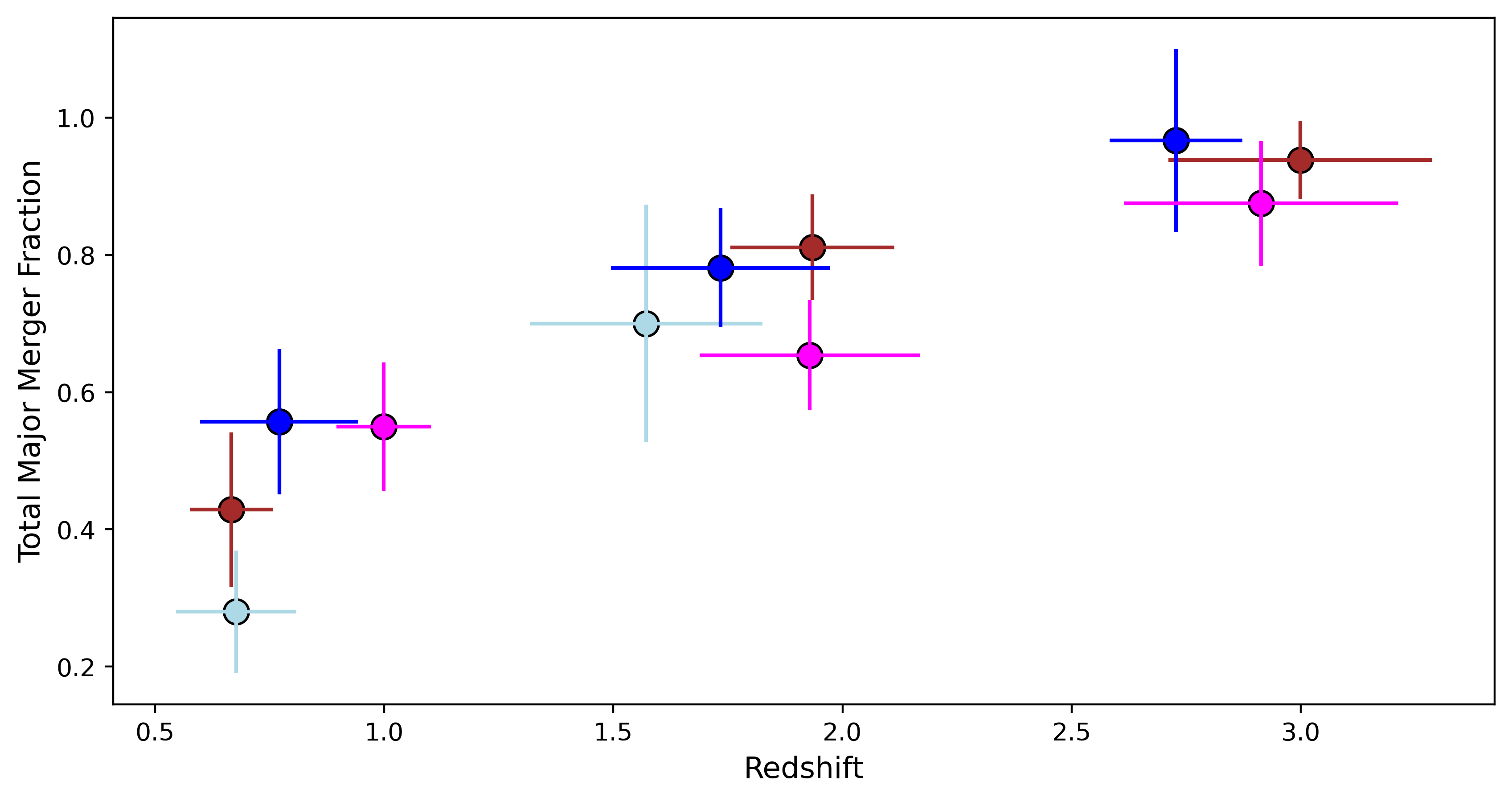}       
\centering
\includegraphics[width=1\linewidth]{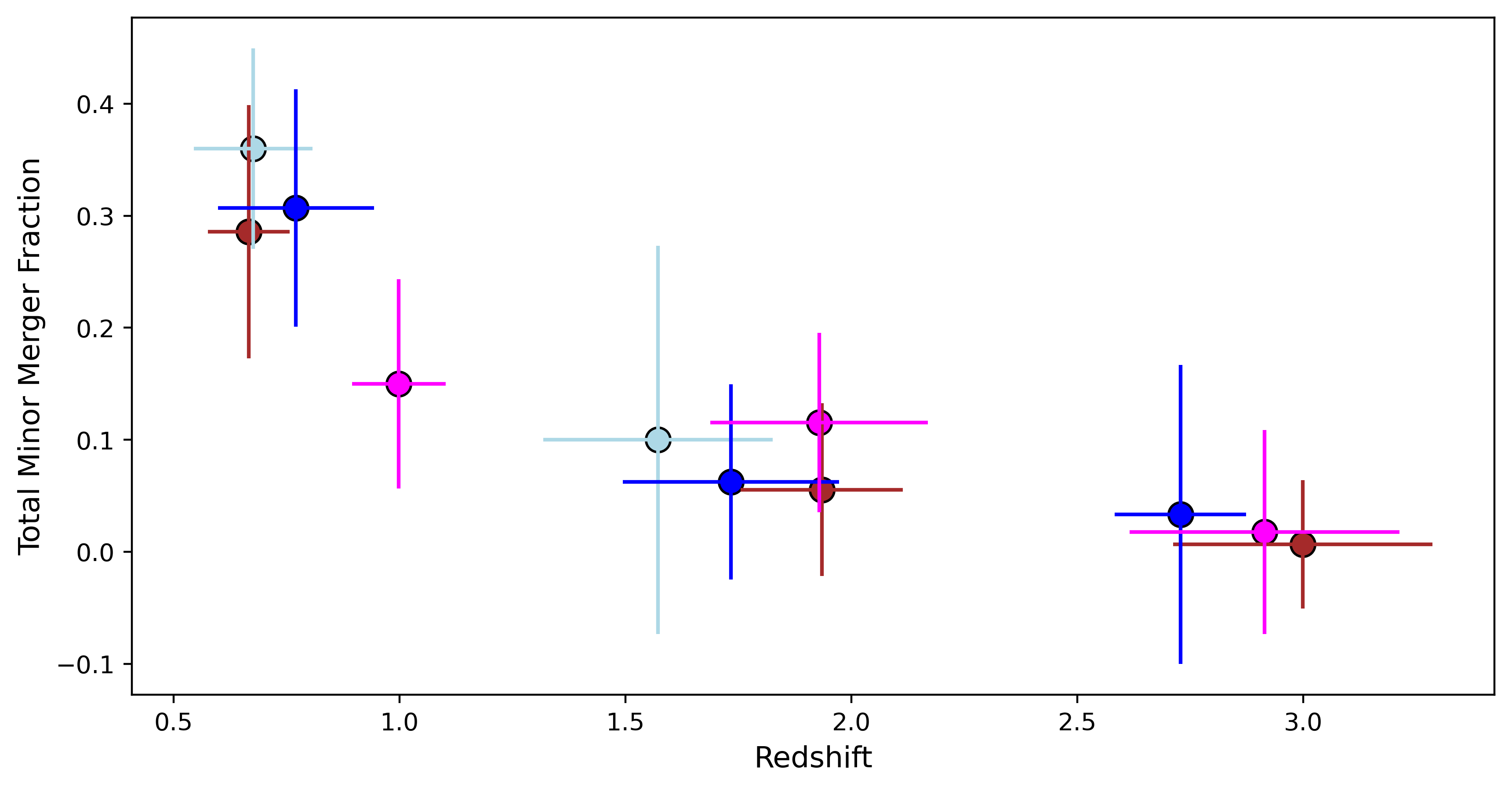}       
\centering
\includegraphics[width=1\linewidth]{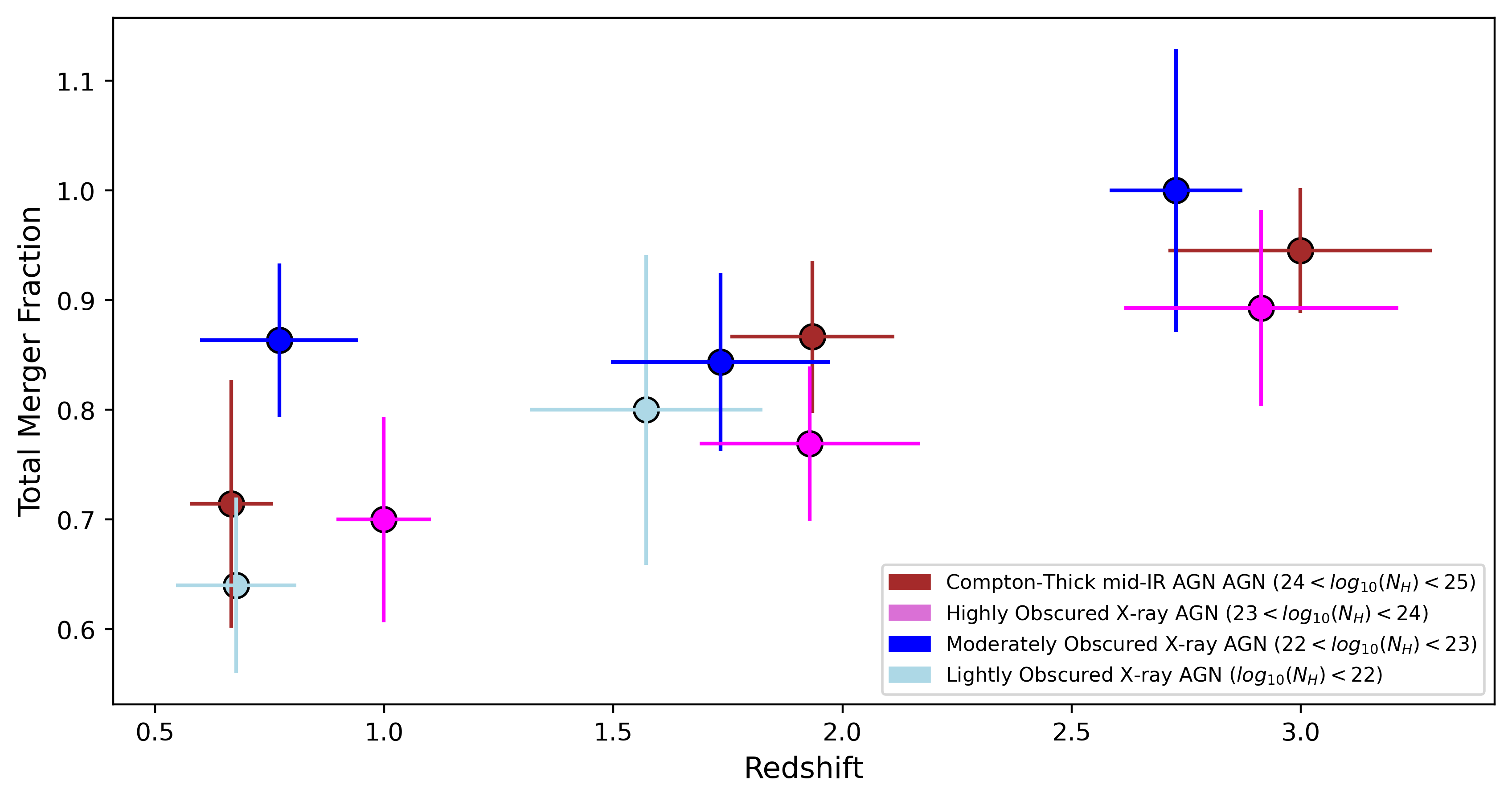}    
\centering
\caption{The major, minor, and total merger fractions resulting from the N23 LDA analysis of our primary AGN sample as a function of redshift (\(x\)-axis) and $N_H$ (point color, simulated for CT mid-IR AGN show in dark red). We observe a rising (decreasing) fraction of major (minor) merger candidates towards higher redshifts; and an increasing total merger fraction with redshift, given that the majority of the merger candidates are classified as major mergers. Here we can see that the majority of the AGN sample exhibit the morphologies characteristic of a merger. (Not shown in the figure are the $8\%$ of the sample classified as non-mergers, $6.8\%$ with degenerate major/minor classifications, and $8\%$ unidentifiable due to poor S$\acute{e}$rsic profile fits.)}
\label{fig:LDA2}
\end{figure}

In our analysis, to qualify as a major-merger candidate, the probability value (\(p\)-value, $p_{\text {merg }}=1 /\left(1+e^{-L D 1}\right)$) resulting from the LDA classification for major mergers was required to be greater than 0.5, while the \(p\)-value associated with the minor merger classification for the same galaxy had to be less than 0.5 (and vice versa to classify a galaxy as a minor merger); see Figure \ref{fig:LDA1} for examples. The relatively small fraction ($6.8\%$ of galaxies with degenerate merger type classifications, i.e. those with LDA values suggesting a high probability of being both major and minor (\(p\)-values greater than 0.5 in both cases), were simply labeled as a ``merger" to indicate an inability to distinguish the merger type (these are included in the total merger fractions plotted in the top panel of Figure \ref{fig:LDA2}). Furthermore, despite the fact that we excluded from our final analysis the N23 LDA classifications for merger stage, we confirmed in Paper I that nearly all AGN in our sample are isolated, as indicated by a lack of a companion galaxy within 50 kpc at the same redshift that would indicate a pre-coalescent merger stage. Therefore, we assume that the merger candidates we identify to be in the post-coalescent stage, minus the handful observed to be in the late merger stage with two visible nuclei.

The results of our LDA analysis are shown in Figure ~\ref{fig:LDA2}, where we see the fraction of AGN in minor and major mergers as a function of both redshift and $N_H$ (where $N_H$ has been simulated for the CT mid-IR AGN as described in Section 2.1.2 of Paper I). From this it is clear that the host galaxy disturbances in a majority of the AGN sample are likely to have been induced by a recent major merger, with an increasing major (decreasing minor) merger fraction observed towards higher redshift.

\begin{figure}
\centering
\includegraphics[width=1\linewidth]{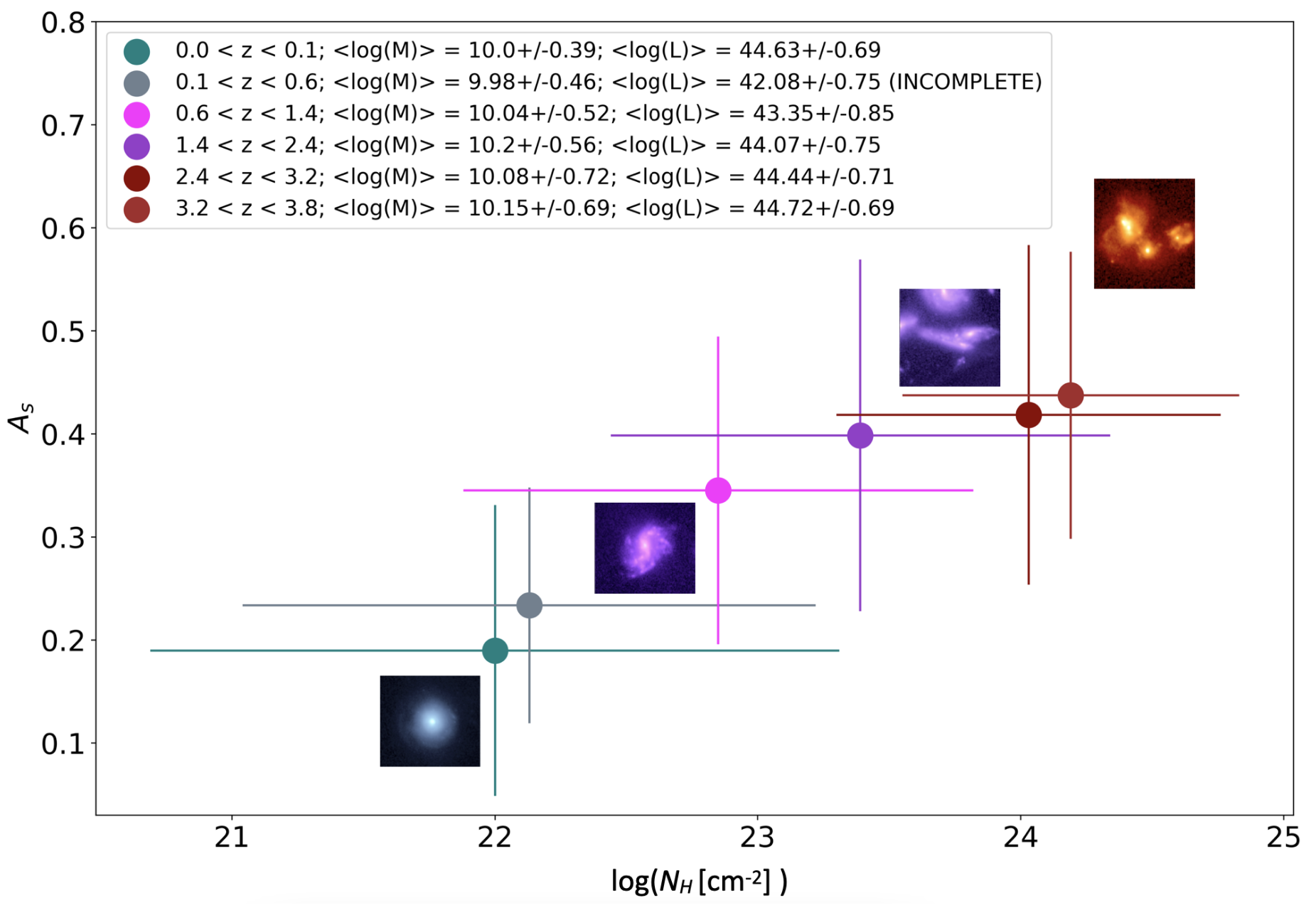}       
\caption{From Paper I \citep{Bonaventura2025}, the trend in the shape asymmetry parameter $A_S$ with $N_H$ and redshift, where the primary AGN sample considered here is contained in the points for $0.6<z<3.8$. Here it can be seen that at Cosmic Noon, the majority of AGN are significantly spatially disturbed; and that at the highest levels of $N_H$ and redshift, the observation of spatial disorder (i.e., an irregular morphology type of a single galaxy or an obvious merger with multiple visible nuclei) in combination with high $A_S$ indicates the early stages of a chaotic merger.}  
\label{fig:AStrend}      
\end{figure}

The results of multi-metric morphology analysis therefore agree with the $A_S$ analysis from Paper I, indicating that the strong spatial asymmetries characterizing the majority of the sample are likely to be the result of major mergers, and that these are common amongst obscured Seyferts at Cosmic Noon. The multi-metric analysis also appears to confirm our speculation from Paper I that Seyferts are unexpectedly evolving according to the cosmological merger rate like quasars, as evidenced by both the decreasing degree of host disturbance and number of hosts with disturbances observed at lower redshift. Figure ~\ref{fig:AStrend} reveals the main findings of Paper I, that there is a trend in the \textit{type} of disturbance with increasing obscuration, alongside the increase of spatial asymmetry: at the highest levels of $N_H$, the galaxies are both more asymmetric and disordered than at lower levels, i.e., the host galaxies of the most heavily obscured AGN have more chaotic morphologies than the hosts of less obscured and presumably more evolved AGNs. 

It would appear that the widespread major merging occurring amongst Seyferts at Cosmic Noon ought to play a significant role in their triggering, while passive/secular processes begin to play an increasingly important role at lower redshifts (where the Seyfert characteristics were originally defined, and perhaps the main reason this class of AGN was not believed to be involved in mergers until now). The control sample morphology analysis, presented in Section 4, lends further support to these arguments.

\begin{figure}
\centering
\includegraphics[width=1\linewidth]{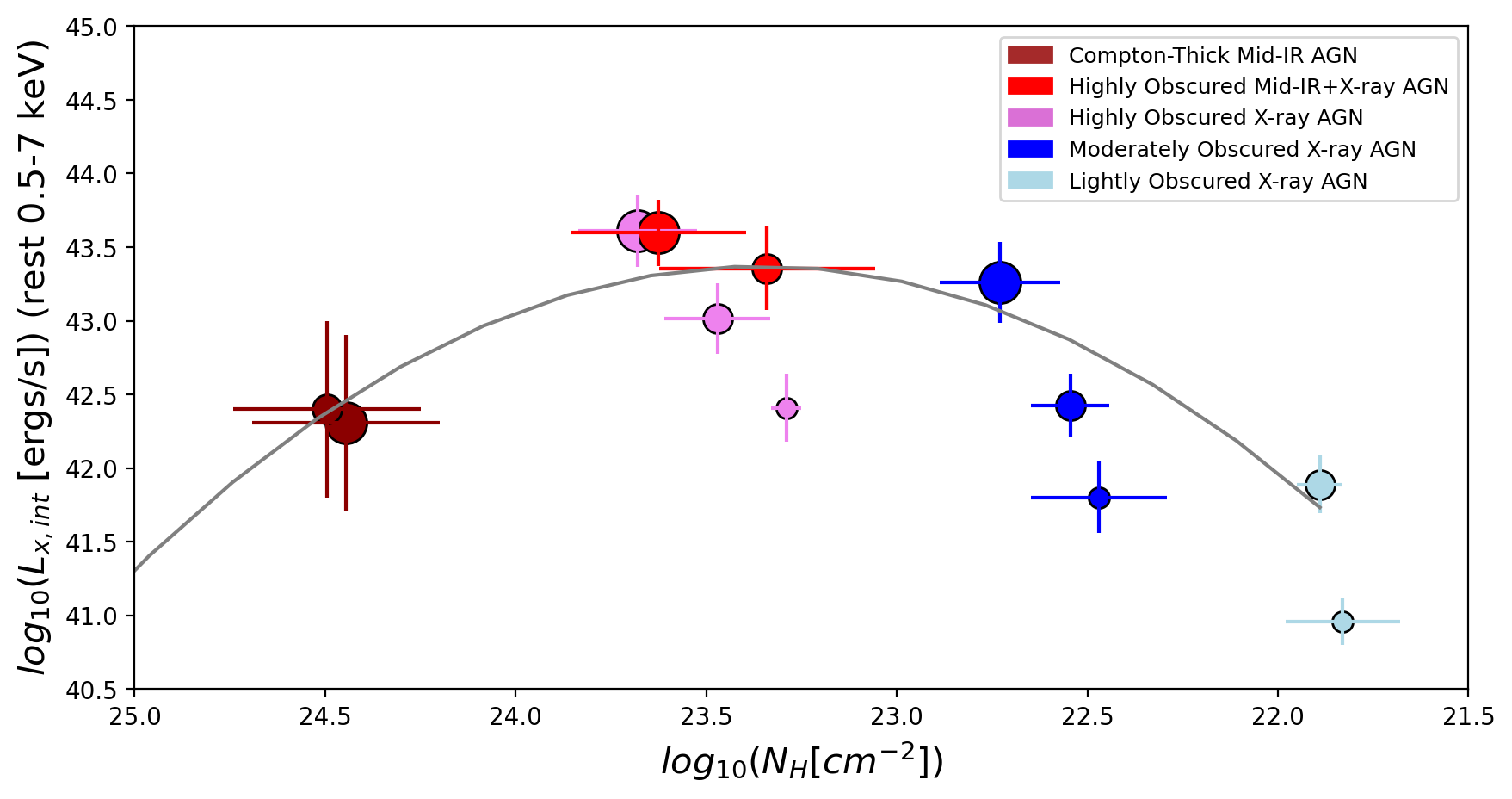}       
\centering
\includegraphics[width=1\linewidth]{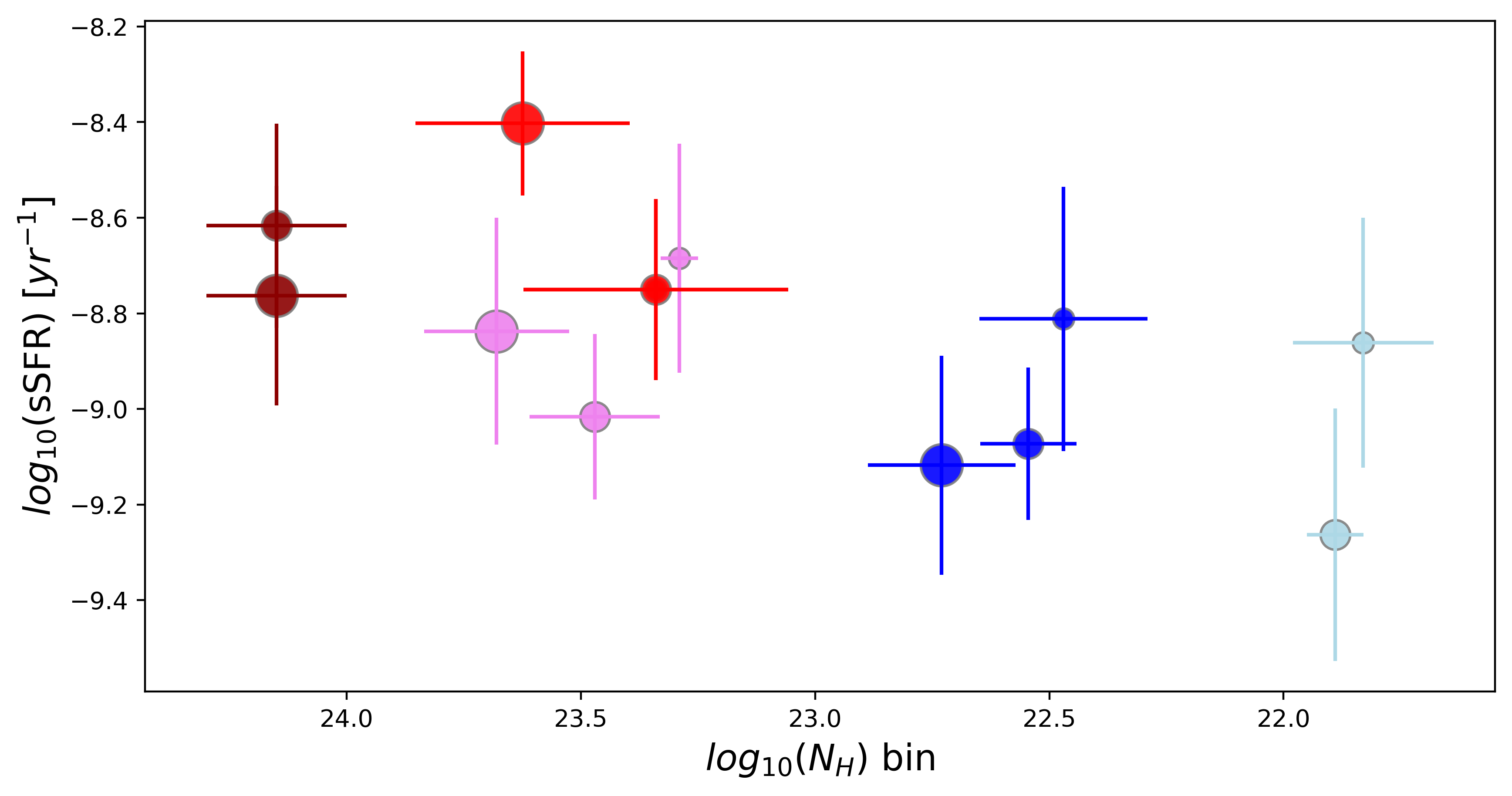}
\caption{Median values of $L_{x,int}$ and mass-normalized star formation rate (specific star formation rate, sSFR) versus $N_H$ bin for the CT mid-IR/X-ray-faint AGN, X-ray/mid-IR-faint AGN, and small subset of AGN detected at both mid-IR and X-ray wavelengths, in multiple redshift bins (the $L_{x,int}$ values for the mid-IR AGN represent the stacked values calculated as described in Section 4.1.1). Symbol sizes increase with redshift, where the smallest to largest points represent the redshift bins 0.5-1.2, 1.2-2.4, and 2.4-3.8. Error bars represent the Median Absolute Deviation of values about the median value in each data bin.  As discussed in Section 4.1.2, the trend plotted in the top figure, tracing the points of the two highest redshift bins, correlates with the trends of decreasing shape asymmetry and $N_H$, and therefore suggests a major-merger-driven Seyfert evolutionary sequence. 
}
\label{fig:Lint}      
\end{figure}

\subsection{Comparison of SED-derived AGN Host Galaxy Properties}

To achieve the second goal of our follow-up analysis to Paper I, we compare the average SED-derived specific star-formation rates and intrinsic X-ray luminosities of the X-ray-bright/mid-IR-faint and X-ray-faint/mid-IR-bright AGN subsamples to check for trends that would either support or contradict the implication that they are linked to a common evolutionary scenario. Given the faintness of the mid-IR AGN in X-rays, this portion of the analysis involved stacking the X-ray fluxes of the mid-IR AGN to a) confirm their CT status that we only presumed in Paper I, proving they are in fact an extension of the X-ray-detected sample into the CT regime (given their otherwise similar host-galaxy properties); and b) to enable a direct comparison of their estimated SMBH accretion rates to their X-ray counterparts via their stacked $L_{x,int}$ (rest 0.5-7 keV) luminosities. 

\subsubsection{Stacked X-ray Luminosities of the mid-IR AGN}

For many of the most heavily obscured AGN, there has been little direct confirmation that they are truly AGN. While the mid-IR AGN sample was identified as such through a broadband SED analysis composed of HST and JWST data, we address that issue by showing that, as a group, they also have the expected X-ray properties for that designation (this issue is also being addressed in a forthcoming paper, G. Rieke et al, submitted to ApJ). This also enabled a comparison of the intrinsic X-ray luminosities and implied accretion rates of the X-ray AGN to the CT mid-IR/X-ray-faint AGN.

To generate stacked X-ray flux measurements of the CT mid-IR AGN, we utilized the CSTACK \citep{Miyaji2008} software, which performs aperture photometry on an input list of sources, using archival Chandra ACIS-I observations (Cycle 15) in each of the soft (0.5-2 keV) and hard (2-8 keV) observed X-ray bands. For our analysis, we chose the deepest available Chandra data in CSTACK, namely the Chandra Deep Field South (CDFS) 7-Mega-Second survey, and stacked the mid-IR AGN in our GOODS-S sample in two redshift bins, $z=1.2-2.4$ (19 AGN) and $z=2.4.-3.8$ (68 AGN). The choice to explore the X-ray properties of the mid-IR AGN as a function of redshift was motivated by the thought that, if we are observing a previously unnoticed, fundamental characteristic of Seyfert evolution at Cosmic Noon, it should be observed consistently throughout the redshift bins of our analysis at $z>\sim~0.5$.

The stacked X-ray fluxes were calculated using the default/recommended CSTACK settings, where the circular apertures used to sum source counts correspond to an enclosed counts fraction of 0.9; and the background level in each image is estimated using a $30''\times30''$ square region centered on the source position, excluding the emission within a 7''-radius around the source. Furthermore, to achieve the most accurate stacked fluxes, we allowed the software to employ its default criteria for automatically rejecting from the input source list those whose apertures were contaminated by flux from nearby resolved sources, as well as the sources that were themselves resolved and heavily influencing the stacked source count. The stacked count rate value calculated by CTSTACK represents a weighted and normalized mean of the rates for the accepted objects, with a corresponding uncertainty value calculated using a bootstrap re-sampling analysis, which is considered to be more accurate than simply using photon-counting statistics (see the CSTACK reference for further details). This procedure resulted in a significant stacked X-ray count rate in each redshift bin of the mid-IR AGN sample, in one or both of the soft and hard X-ray bands, confirming their status as the most heavily obscured, CT AGN of our full AGN sample. 

To estimate the intrinsic, rest-frame $0.5-7$ keV X-ray luminosity ($L_{x,int}$) corresponding to the stacked X-ray count rate generated by CSTACK in the soft and hard X-ray bands, per redshift bin, we first calculated the corresponding physical flux value using the PIMMS $v4.14$ tool available in the Chandra Proposal Planning Toolkit. The PIMMS tool translates an input X-ray count rate, flux, or flux density value in a given energy range, specific to a chosen Chandra cycle and detector/grating/filter combination, into an output count rate, flux, or flux density value in the desired output energy range, using the input model parameters that describe the source emission. In other words, PIMMS interpolates the spectral model to the input parameters, folds it through (i.e., multiplies) the Chandra effective area curves, and integrates over the user-specified input energy band(s) to determine the appropriate conversion factors.

To model the CT AGN in PIMMS, we chose an absorbed power-law with a photon index of 1.8; a Galactic $N_H$ absorption value of $6.8 \times 10^{19}$ cm$^{-2}$, (obtained from the online HEASoft tool for the GOODS-S coordinates); and adopted a redshifted $log_{10}(N_H/cm^{-2})$ value of 24.5 to represent the typical observed value for $\sim90\%$ of Compton-thick AGN, where $24<log_{10}(N_H)<25$ brackets the minimum and maximum observed values \citep[e.g.][]{Levenson2006, Levenson2014, Kammoun2019}. For the input energy range, we used the observed X-ray flux range corresponding to the soft or hard band used in CSTACK, and entered the desired rest-frame energy range of $0.5-7$ keV for the output energy.  With the resulting stacked flux values from PIMMS, we converted to $L_{x,int}$ using the luminosity distance corresponding to the average redshift of each redshift bin, and the k-correction value calculated using the Chandra CIAO tool \textit{calc\_kcorr} (https://cxc.cfa.harvard.edu/sherpa/ahelp/calc\_kcorr.html) at the desired redshift and input/output energy range, via the following relation: $L_X=F_X \times 4 \pi d_L^2 \times k(z)$. 

Given that the CSTACK soft-band count rates are likely to be contaminated by host-galaxy emission, and/or a soft X-ray excess from the AGN, and/or extreme absorption, we chose for our final analysis to use only the stacked X-ray luminosities derived from the CSTACK hard-band count rates. 
This ensured that the stacked X-ray luminosities were being normalized by the AGN emission longward of $\sim4$ keV that should be minimally affected by these processes and therefore most accurately reflect the true intrinsic power-law emission of the AGN (see the AGN SED components in Figure 1 of \citet{Hickox2018}, and the SEDs of minimally to highly obscured AGN in Figure 4 of that work, with a varying level of AGN dominance over the host galaxy). Furthermore, the stacked X-ray luminosities were calculated in a way that is consistent with the X-ray luminosities \citep{Lyu2022} and associated $N_H$ estimates \citep{Luo2017} for the X-ray AGN sample to which they are compared in Figure \ref{fig:Lint}: we adopt the same intrinsic power law model and de-absorption assumptions (photoelectric and Thomson absorption only) to arrive at the intrinsic X-ray luminosity, with no scaling by a reflection efficiency.

\subsubsection{X-ray luminosities and Specific Star Formation Rates}

In Figure~\ref{fig:Lint} we compare the average intrinsic X-ray luminosity, $L_{x,int}$, of the mid-IR and X-ray AGN subsamples in different bins of line-of-sight $N_H$ obscuration. At the highest redshifts ($z>\sim1.2$), we observe $L_{x,int}$ peaking amongst the highly obscured X-ray AGN subset; and the the stacked $L_{x,int}$ for the mid-IR CT AGN lying nearly an order of magnitude lower. We also observe an unmistakable trend of the average $L_{x,int}$ steadily decreasing through the X-ray AGN subsets with coincidentally decreasing obscuration (and also spatial asymmetry, $A_S$, as presented in Paper I). This picture mirrors the simulation results of \citet{Mcalpine2020} who show that, at $z>1$, $\sim50\%$ of BH activity triggered by major mergers occurs within a dynamical time \(after\) the active merging phase, in the post-coalescent remnant. Furthermore, they found that galaxies with the stellar ($M_* \sim 10^{10} M_\odot$) and black-hole masses ($M_* <\sim 10^7 M_\odot$) typical of our sample show the greatest enhancement of BH activity from major mergers.

Similarly, the median mass-normalized star-formation rates (specific star formation rate, sSFR) we measure track the expected trend of this property as a function of a major-merger timeline (e.g., as shown in the merger simulations of \citet{Sotillo2022, Schechter2025}): we observe the peak average sSFR coinciding with the most highly obscured X-ray AGN subset, along with $L_{x,int}$, which bolsters our view that this may be a distinct Seyfert AGN phase marking the period of maximal accretion. While star formation in a sub-quasar AGN host-galaxy can exist largely independently of the central BH activity, major-merger simulations of galaxies with stellar masses and morphologies similar to those characterizing our sample show that bursts of star formation predictably occur during pericentric passages prior to coaleascence, and also with a delay after coalesecence around the same time that the BH accretion rate also maximizes \citep{Ellison2025, Schechter2025}. This would indicate that the coincident peak of both $L_{x,int}$ and SFR marks the window of time in which the predicted simultaneous bursts of AGN luminosity and star-formation occur just prior to the AGN phase where feedback begins to counter the star formation activity and lower the measured SFR.

To complete the connection among these phases, \citet{Hickox2009} found that \(Spitzer\)/IRAC-detected infrared AGN lie mostly in the `blue cloud' while \(Chandra\)-detected X-ray AGN are more concentrated in the `green valley' with redder colors and a modest tendency toward higher optical luminosities and hence larger masses. These authors extensively compared and contrasted X-ray AGN and mid-IR AGN to reveal many similarities in their host galaxy properties in addition to their stellar colors, such as their clustering properties, suggesting they belong to the same population. This leads us to conclude that the CT-obscured mid-IR AGN and larger subsample of lightly-to-highly obscured X-ray AGN are not distinct AGN classes but related through major-merger evolution.\footnote{We note that comparisons with previous work \citep[e.g.,][]{George2023} must be done cautiously, since (1) this comparison could be affected by the presence of X-ray AGN in the red sequence, a population absent from the CT MIRI sample \citep{Hickox2009}; and (2) studies defining ``obscured'' through X-rays \citep[e.g.,][]{Ricci2017, George2023} miss the highest $N_H$ bin entirely.}

Taken together, the average SED properties of the mid-IR and X-ray AGN samples suggest we are statistically likely to be witnessing the AGN emission footprint that characterizes an evolving major merger. Further support is lent to this interpretation from the observed features of the small subset of AGN with both X-ray and mid-IR emission, detected at $z > 1.2$. These systems exhibit similar peak values of $L_{x,int}$ to the most highly obscured X-ray-bright/mid-IR-faint AGN, and have similar $N_H$ levels but that extend to slightly lower levels. Given the minimal overlap between the X-ray and mid-IR AGN subsamples, meaning most of the X-ray AGN are undetected in the mid-IR and vice versa, this small fraction of AGN with both X-ray and mid-IR emission appears to represent a short-lived transitional phase between the two. 

\subsection{Comparison of AGN Trends with Inactive Control Sample}

\begin{figure}
\centering
\includegraphics[width=1\linewidth]{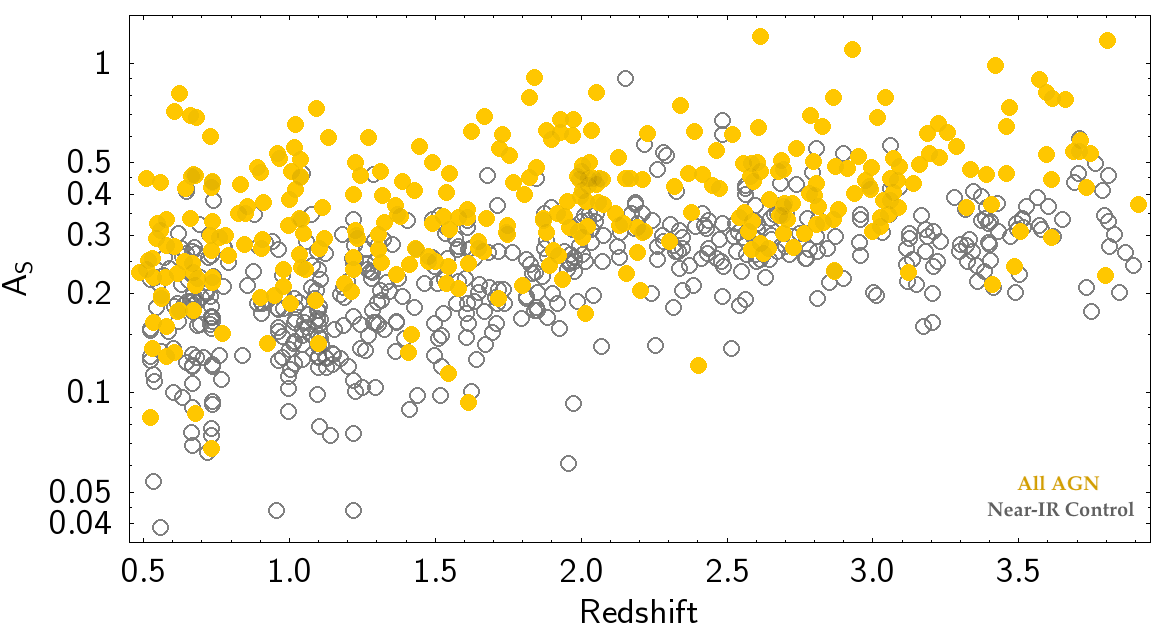}  
\centering
\includegraphics[width=1\linewidth]{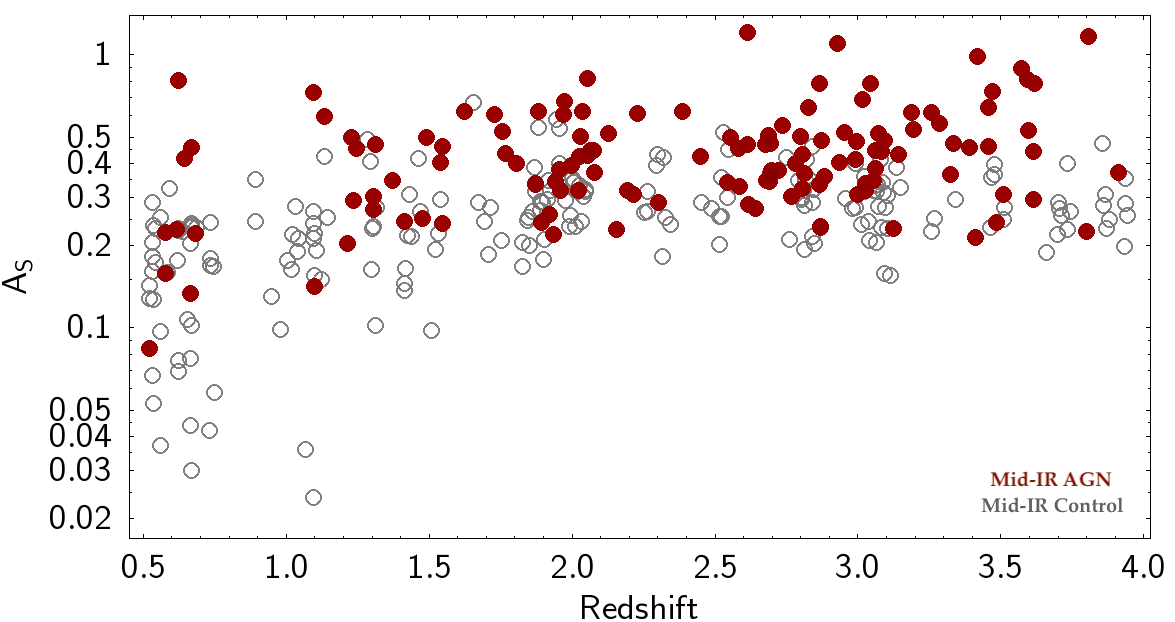}       
\centering
\includegraphics[width=1\linewidth]{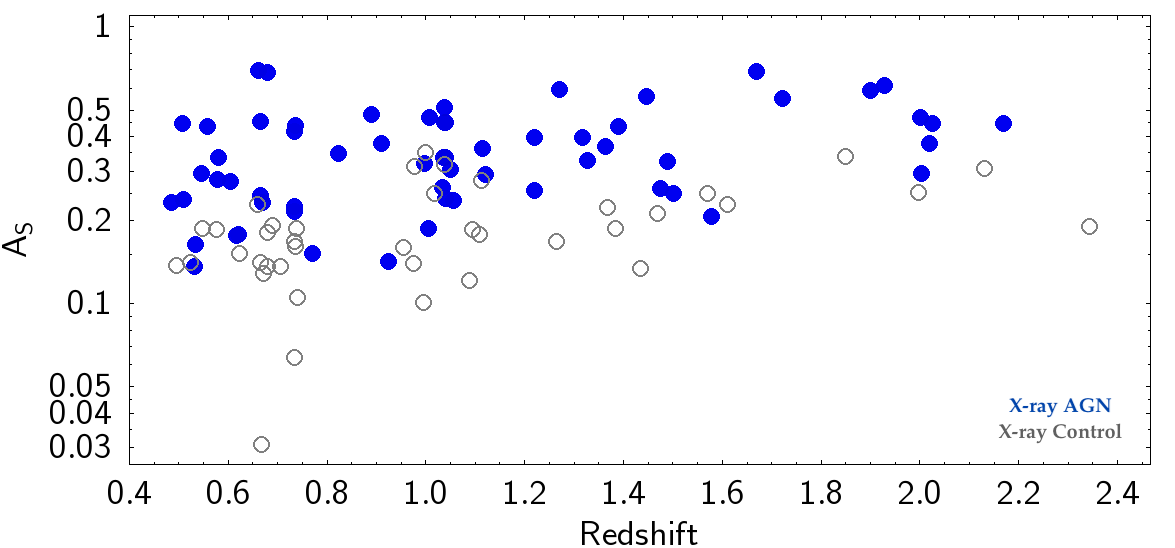}      
\caption{A comparison of the shape asymmetry distributions of the AGN subsamples and corresponding matched control samples. The top panel of the figure shows our full primary AGN sample (yellow, $66\%$ with $A_S>=0.3$) compared to the near-IR-selected control sample ($25\%$ with $A_S>=0.3$); the middle panel shows the mid-IR AGN (red, $79\%$ with $A_S>=0.3$) and corresponding MIRI-selected control sample ($29\%$ with $A_S>=0.3$); and the bottom panel shows the X-ray AGN (blue, $60\%$ with $A_S>=0.3$) compared to the X-ray-selected inactive control sample ($14\%$ with $A_S>=0.3$). Here it can be seen that the majority of the property-matched inactive control galaxies exhibit $A_S$ values largely confined to less than 0.35 (matching the $A_S$ values and range of simulated non-mergers), while a significant fraction of the AGN extend to higher asymmetry values.}    
\label{fig:assym} 
\end{figure}

\subsubsection{Control Sample Morphologies}

We measured the spatial morphologies of the near-IR, mid-IR-, and X-ray-selected control samples using the same methodologies described in detail in Paper I, namely through human visual inspection and computer vision with the \(statmorph\) software. The former method involved qualitatively classifying each galaxy into one of the following morphological classes: point source, spheroid, disk (either symmetric/ordered, or mildly or strongly asymmetric/disturbed), irregular, and merger. For the latter method, we again adopted a conservative definition of $As\ge0.3$ to define a strong host-galaxy disturbance, whereas \citet{Pawlik2016}, who introduce $A_S$, define a strong disturbance as $As\ge0.2$. We also split the AGN and control samples into two mass bins, $10^9$-$10^{10}$ $M_{\odot}$ and $10^{10}$-$10^{11}$ $M_{\odot}$, to observe any morphological trends with this property.

The results of the morphological analysis of the near-IR control sample revealed, for stellar masses between $10^{10}$ and $10^{11}$ $M_{\odot}$, $9.4\%$ strongly disturbed, as compared to $61\%$ of AGN in the same mass range. The numbers come closer together for the lower-mass bin of $10^9$-$10^{10}$ $M_{\odot}$, meaning more disturbance is observed amongst the controls: $33\%$ of the controls in this mass bin are strongly disturbed, compared with $70\%$ of AGN. For the MIRI-selected control sample, a similar trend is observed: $84\%$ of the CT mid-IR AGN exhibit strong spatial asymmetries at stellar masses between $10^9$-$10^{10}$ $M_{\odot}$, compared to $35\%$ for the MIRI-selected control galaxies; while for the stellar mass bin $10^9$-$10^{10}$ $M_{\odot}$, $59\%$ of the AGN are highly disturbed and only $5\%$ of the controls. 

Overall, $29\%$ of the MIRI inactive controls, $14\%$ of the X-ray inactive controls, and $25\%$ of the near-IR control sample are highly disturbed. A two-proportion \(z\)-test applied to each AGN sample and matched control shows, with strong confidence, that they are unlikely to be drawn from the same population, based on the following results: \(z=8.97, p=2.9 \times 10^{-19}\) for the mid-IR AGN and matched control sample, with a difference in proportions of 0.49 ($95\%$ $\mathrm{Cl}$ $0.40-0.58$); \(z=5.61, p=2.0 \times 10^{-8}\) for the X-ray AGN and matched control sample, with a difference in proportions of 0.464 ($95\%$ $\mathrm{Cl}$ $0.302- 0.626$); and \(z=11.24, p=2.7 \times 10^{-29}\) for the full AGN sample and matched near-IR control sample, with a difference in proportions of 0.36 ( $95\%$ $\mathrm{Cl}$ $0.3-0.4$)

As for morphological type, we observe that both the X-ray- and mid-IR-matched inactive control sample galaxies are mostly disks, mirroring the findings for the X-ray AGN, but different than the highly disturbed and irregularly shaped mid-IR AGN (see Figure~\ref{fig:AStrend} and Paper I). From this it is clear that the high degree and incidence of disordered morphologies characterizing the mid-IR AGN is unique to their nature as AGN (noting that the mid-IR control galaxies show a slightly higher fraction of irregular galaxies than the X-ray-selected control galaxies).
 
A comparison of the shape asymmetry values of all inactive controls to the simulated merging and non-merging galaxies in N19 shows that they lie within the same approximate range as the non-merging galaxies, all confined mostly to $0.1<A_S<0.3$. The simulated merging galaxies in that study and our AGN merger candidates overlap this $A_S$ range but also significantly extend beyond to higher asymmetry values. If we take a conservative approach and consider all AGN in our sample with $A_S<0.3$ as non-mergers, the results of Paper I remain, that a large fraction of our AGN sample are statistically highly disturbed. The study of the control sample morphologies presented here lends further support to our conclusion that, statistically, AGN are more likely to be involved in mergers than inactive galaxies.

\section{Discussion}

Our study spanning Papers I and II recovered and expanded upon the tentative result of \citet{Donley2018}that mid-IR-detected/X-ray-undetected AGN significantly reside in irregular, highly disturbed hosts appearing as merger remnants. Furthermore, it corroborated, and extended to a higher redshift and level of $N_H$, the results of \citet{Kocevski2015}, that a significant fraction of obscured X-ray-selected AGN at Cosmic Noon exhibit the galaxy-scale spatial disturbances associated with merging activity, and that most have a disk morphology. 

Our result is also in agreement with \citet{Ellison2025}, who show that AGN incidence reaches a maximum shortly {\it after} coalescence, not in the pre-coalescent or early merging stage where a pair of nuclei is still visible. We note that their result is based on a number of AGN metrics, but does not include the heavily obscured stage, whereas our study tracks this trend to its logical end point in the CT mid-IR AGN subsample. Finally, our finding that mergers are associated with dust-obscured AGN is in agreement with the recent paper by \citet{Lamarca2024} and the trend found by \citet{Dougherty2024}.

\subsection{Disk Morphologies of Merger Remnants}

Until the release of Paper I, mergers were not considered to be a possible cause of disturbed disk morphology in AGN \textit{after} coalescence, only before (e.g., \citealp{Kocevski2015}). This was based on hallmark early studies on AGN triggering mechanisms that indicated that mergers, particularly major mergers, would not result in a remnant with an ordered disk shape. However, as detailed in Paper I, numerous galaxy merger simulations have shown that disks commonly emerge from gas-rich major mergers, whether or not the interacting galaxies have a disk morphology before the merger. This powerful, relatively new understanding of the physics of galaxy mergers, in combination with our observation of significant signs of merger-induced disturbances in AGN host galaxies owing to the availability of JWST/NIRCam data, strongly supports a scenario in which sub-quasar Seyfert galaxies are evolving similarly to quasars at $z>\sim0.5$ and $9<log(M_*/M_\odot)<11$.

\subsection{AGN Emerge in Early Post-coalescence}

\begin{figure*}
\centering
\includegraphics[width=1\linewidth]{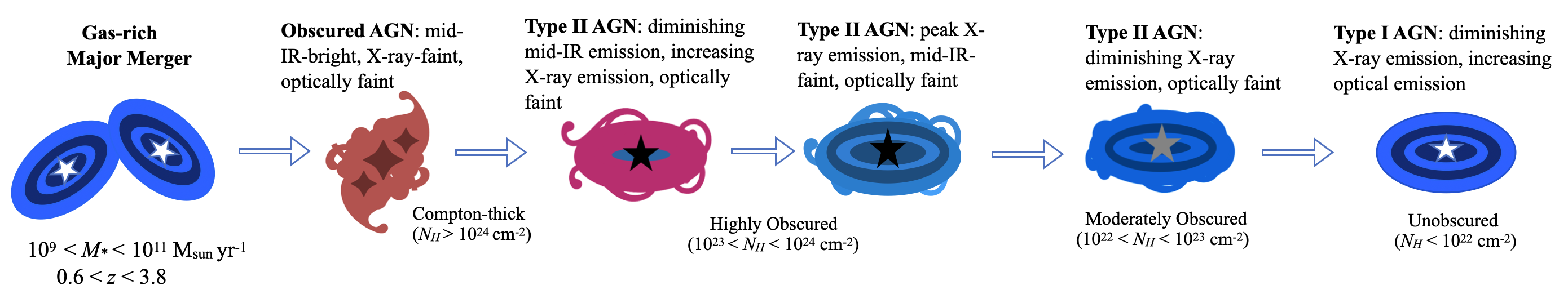}  
\centering
\caption{A second branch of the historically accepted Seyfert evolutionary sequence that allows for merging activity as a triggering mechanism, in addition to solely secular processes. This discovery amongst Cosmic Noon AGN host galaxies with $10^9<M_*/M_{\odot}<10^{11}$ and moderate AGN luminosities was enabled by the significant increase in spatial resolution and photometric sensitivity of JWST/NIRCam over its predecessor instruments, which allowed for the detection of a significant number and degree of merger-induced morphological disturbances.}
\label{fig:cartoon}
\end{figure*}

The X-ray emission of an AGN originates in its central engine, thus X-ray selection is sensitive to a broad range of evolutionary stages, excepting the most heavily obscured one. The infrared emission  associated with the circumnuclear torus of dust and gas is nearly universal and also appears to be intrinsic. However, the infrared output associated with this embedding material fades as it is expelled by powerful winds and outflows from the central engine. As a result, the existence of a substantial population of heavily obscured, infrared-dominated AGN must represent a transitory stage relatively early in the life of the AGN, given that once the obscuring material is dispersed, it is inconceivable that it would be re-assembled. This pins the highly obscured and spatially disordered mid-IR AGN in our sample to the late major-merger phase, given that the only physical process known to induce such a galaxy-wide spatial disruption and redistribution of gas and dust is a major merger. This then leaves two possibilities for the interpretation of the strongly disturbed disk morphologies that we have found in the X-ray obscured AGN: that they are either in the (1) pre- or (2) post-coalescent stage of a major merger. Given that it is not always possible to morphologically distinguish the pre- and post-coalescent phases of a merger due to degeneracies in their morphological properties (see N22 and our related discussion in Section 3), a search around each AGN within $\sim 50 kpc$ for a nearby galaxy detected at the same redshift can achieve this goal. The fact that nearly all of the X-ray AGN in our sample appear isolated by this criterion leads us to conclude that they are largely in a post-coalescent phase. 

This conclusion can be compared with the study of \citet{Ellison2025}, who generated a sample of 21,235 galaxy pairs from the Sloan Digital Sky Survey DR7, 8141 post-merger systems from the Ultraviolet Near Infrared and Optical Northern Survey (UNIONS) \citep{Gwyn2025}, and a large control sample of non-AGN galaxies. They used the Multi-Model Merger Identifier machine vision pipeline to determine a post-merger timeline for the galaxies, and identify AGN through optical emission lines and the WISE W1 - W2 color. They found (see their Figure 4) a modest enhancement in AGN incidence pre-merger, with a much larger enhancement (by a factor of three) just after coalescence, that persists at a lower level for 1 Gyr post-merger. 

The time immediately post-coalescence is when interstellar clouds will have fallen into the potential well of the merged system and nuclear obscuration will be the strongest. Thus, any AGN activated during this period will tend to be embedded or strongly obscured and will be found preferentially in the infrared. This is exactly as in  \citet{Lamarca2024}, who studied AGN and mergers in the Kilo Degree Survey (KiDS; \citealp{dejong2013}); they found an excess of IR AGN by a factor of 2 - 3 in mergers, but only mild excesses for AGN identified through X-rays or SED analysis. 

The infrared data used in these studies was not sufficiently deep and comprehensive to identify deeply embedded AGN, which are the subject of our study. The trends they find would be expected to be stronger for these more extreme infrared-identified AGN, consistent with our results. At the same time, the merger paradigm does not appear to apply to all AGN, as 1) we only observe mergers in our sample with statistical significance at $z>\sim1$; and 2) Figure~\ref{fig:assym} shows that there is significant overlap between the shape asymmetry values for our AGN and control samples in the range of $A_S=0.2-0.3$ (reminding the reader also that the simulated non-merging galaxies of N19 show shape asymmetry values extending up to $A_S=0.3$). That is, although we have found an association of mergers with obscured AGN, the evidence is still consistent with a significant subsample of AGN at $z<1$ being fueled stochastically rather than in major mergers.

\subsection{A Merger-driven Seyfert Evolutionary Scenario}

On the basis of these results, we suggest an evolutionary scenario for the merger-induced branch of nuclear activity, depicted in Figure~\ref{fig:cartoon} in a style similar to the sketches in \citet{Hickox2009}. 
We propose that the CT mid-IR AGN mark the early coalescence phase, where we necessarily expect to observe their descendants in the form of merger remnants, showing an initial increase in X-ray emission as the CT circumnuclear material begins to clear. As the remnant ages, we expect steadily diminishing levels of $N_H$, mid-IR emission, and spatial asymmetry as it reforms a disk shape after the merging event (\citealp{Sotillo2022} show that $58\%$ of simulated major mergers amongst Milky Way- and M31-like galaxies reformed a destroyed disk shortly after the merger). In other words, as the most highly obscured X-ray AGN emerge from the CT mid-IR phase, their host galaxies settle into increasingly ordered and symmetric disk shapes with time past the merger event, where IR-emitting dust becomes increasingly confined to the disk and settles into a conventional circumnuclear torus. We believe this scenario to be the most likely to explain the set of host galaxy properties we measure for the X-ray AGN, as well as the relatively small number that are simultaneously detected in the mid-IR. 

This is in agreement with the physical model of AGN feedback discussed in \citet{Hopkins2006}: due to their smaller, sub-quasar black hole masses, feedback in Seyferts is not expected to reach the most powerful `blowout' phase characterizing quasars; instead, it results in a weaker blast wave that fails to rapidly shut down host-galaxy star formation and result in the spheroid morphology characteristic of quasar remnants. This downsized version of quasar evolution appears to manifest as a population of AGN that (1) maintain, rapidly reform, or newly develop a gaseous disk morphology after a major merger; and (2) continue to form stars and feed the nucleus, with a steadily declining rate of BH accretion and star formation. This is how our X-ray AGN sample behaves.

In our comprehensive study of CT mid-IR and lightly-to-highly obscured X-ray AGN in Papers I and II, we have shown that these two apparently distinct types of AGN actually belong to the same AGN class, based on the observed morphological and AGN emission pattern appearing to connect them in a merger-driven evolutionary sequence. An additional piece of evidence that bolsters this conclusion is the finding that the dominant morphology of the mid-IR inactive control sample (mostly disks) differs from the matched mid-IR AGN sample (mostly irregular). In other words, if a high level of spatial disorder had been found to characterize both the non-AGN and AGN mid-IR-detected galaxies in our samples, then it would not be possible to attribute the highly disturbed and disordered morphologies of the mid-IR AGN to their nature as AGN. Likewise, the significant levels of disk disturbance observed in the X-ray AGN as compared to their non-AGN X-ray disk counterparts, allows us to firmly conclude that their morphologies are a result of their nature as merger-induced AGN.

Finally, it is remarkable, given our large AGN sample size that draws from a $\sim7$-Gyr span of cosmic history, that we do not witness a wide array of uncorrelated morphological and emission properties, but a narrow set that aligns with that expected from an AGN evolving in a major merger. In other words, each subsample of AGN is characterized by a set of correlated, not random, properties; and the overall set of properties of each subsample appears tied to that of the other subsample according to a predictable evolutionary trend.

\section{Conclusions}

Given the comprehensive morphological analysis of a large and representative sample of Seyferts we present in Paper I and here in Paper II, we conclude that the triggering of sub-quasar AGN at Cosmic Noon must in part be due to widespread merging, in tension with previously held beliefs that they only passively evolve. By comparing the morphologies and emission properties of  CT mid-IR/X-ray-faint and X-ray-bright/mid-IR-faint AGN subsamples, we find that they represent distinct phases of a merger-driven evolutionary sequence, as opposed to distinct AGN classes. The various lines of evidence in support of this theory have been laid out in Paper I and the current paper, and is most convincing when comparing the AGN host-galaxy properties to an inactive control sample presented here in Paper II: the pattern of morphological disturbances and types observed amongst the AGN sample is not found in the large sample of inactive controls. Therefore, we are left to conclude that the high level of disturbance and irregularity of the mid-IR AGN is due to their being late-stage mergers; and likewise, that the high fraction of disturbed disk morphologies observed amongst the X-ray AGN translates to them being the post-merger remnants with steadily decreasing X-ray luminosities and levels of obscuration as they evolve in time past the merger event.

\vspace{5mm}

\begin{acknowledgments}
\section{Acknowledgments}
This work was supported by NASA grants NNX13AD82G,  1255094, and the NASA contract for NIRCam to the University of Arizona, NAS5-02015. This work is based on observations made with the NASA/ESA/CSA James Webb Space Telescope. SA, NB, DJE, KH, ZJ, BR, MR, and CNAW acknowledge support from the NIRCam Science Team contract to the University of Arizona, NAS5-02015.  DJE is further supported as a Simons Investigator. SA, JL, IS, and GHR acknowledge support from the JWST Mid-Infrared Instrument (MIRI) Science Team Lead, grant 80NSSC18K0555, from NASA Goddard Space Flight Center to the University of Arizona. These observations are associated with JWST GTO programs \#1180, \#1181, and \#1207, and GO program \#1963. AJB acknowledges funding from the ``FirstGalaxies" Advanced Grant from the European Research Council (ERC) under the European Union's Horizon 2020 research and innovation program (Grant agreement No. 789056). The work of CCW is supported by NOIRLab, which is managed by the Association of Universities for Research in Astronomy (AURA) under a cooperative agreement with the National Science Foundation. WB and FDE acknowledge support by the Science and Technology Facilities Council (STFC) and by the ERC through Advanced Grant 695671 "QUENCH".
\end{acknowledgments}
\facility{JWST, Chandra}

\bibliographystyle{aasjournal}

\end{document}